\documentstyle[12pt,epsfig]{article}
\let\section=\subsection     \let\subsection=\subsubsection                

\def\Journal#1#2#3#4{{#1} {\bf #2} (#4) #3}
\def\AP{\em Ann. Phys.}

\def\NPA{{\em Nucl. Phys.} A}
\def\NPB{{\em Nucl. Phys.} B}

\def\PLB{{\em Phys. Lett.} B}

\def\PRL{\em Phys. Rev. Lett.}

\def\PRD{{\em Phys. Rev.} D}
\def\PRC{{\em Phys. Rev.} C}

\def\ZP{\em Z. Phys.}

\begin{document}
\begin{center}
   {\large \bf FROM MESON-NUCLEON SCATTERING TO}\\[2mm]
   {\large \bf VECTOR MESONS IN NUCLEAR MATTER}\\[5mm]
   B.~FRIMAN$^{a,b}$, M.~LUTZ$^a$ and G.~WOLF$^c$\\[5mm]
   {\small \it $^a$Gesellschaft f\"ur Schwerionenforschung (GSI) \\
   Postfach 110552, D-64220 Darmstadt, Germany \\
   $^b$Inst. f\"ur Kernphysik, TU Darmstadt \\
   D-64289 Darmstadt, Germany\\
   $^b$KFKI, RMKI, H-1525 Budapest, POB.~49, Hungary \\[8mm] }
\end{center}

\begin{abstract}\noindent
We employ meson-nucleon scattering data to deduce the properties of
the low-mass vector mesons in nuclear matter, and present results
for the $\rho$ and $\omega$ in-medium spectral functions. The
corresponding thermal emission rate for lepton pairs is also
discussed.
\end{abstract}

\section{Introduction}

The in-medium properties of hadrons is a topic of high current
interest. This is manifested e.g. by the flurry of activity
triggered by the observed enhancement in the production of low-mass
lepton pairs~\cite{CERES,misko,Helios3} in relativistic
nucleus-nucleus collisions. While ``standard'' models, where the
properties of the hadrons are not modified in the medium, fail to
repoduce the data~\cite{Drees}, models where medium effects of some
kind are invoked, are more successful. The low-mass enhancement in
the lepton-pair spectrum has been interpreted in terms of a
modification of the mass and/or width of the $\rho$ meson in a
dense hadronic environment ~\cite{likobrown,BR,RCW}.

In this talk we present a novel approach, which allows us to
systematically explore the properties of $\rho$ and $\omega$ mesons
in nuclear matter. We employ a general scheme, the low-density
expansion, to establish a connection between the meson-nucleon
scattering amplitudes and the properties of mesons in nuclear
matter~\cite{LDT}. This scheme leads to a systematic expansion for
the in-medium self energy of a meson (in general any particle), in
terms of the corresponding vacuum scattering amplitudes. The
vector-meson--nucleon scattering amplitudes are not directly
related to experimental observables. However, it may be possible to
determine them indirectly in a coupled-channel scheme, where the
$\rho N$ and $\omega N$ channels enter in intermediate and final
states of measured processes. Here we present a calculation along
these lines, where we fix the parameters of an effective field
theory by fitting the available meson-nucleon scattering data in
the relevant energy range. We then employ the scattering amplitudes
to construct the vector-meson self energies in nuclear matter to
leading order in density. Our results are relevant for the study of
vector mesons in nuclei and perhaps in nucleus-nucleus collisions
at GSI energies. On the other hand, at CERN energies, where the
medium presumably is drastically different from the cold nucleon
liquid addressed in this calculation, our results should be applied
with caution.

The aim of this work is to obtain the leading in-medium correction
to the properties of the $\rho$ and $\omega$ mesons in a tenable
scheme, where subleading terms can be systematically computed. A
major advantage of such an approach is that it is possible to check
the convergence of the expansion. On the other hand, at present it
does not allow us to address the question whether modifications of
the properties of vector mesons in matter are connected with the
restoration of chiral symmetry in matter. We leave this intriguing
question for future work.

\section{Meson-nucleon scattering}

In this section we describe a relativistic and unitary coupled
channel approach to meson-nucleon scattering~\cite{LWF}. The
following channels are considered: $\pi N$, $\rho N$, $\omega N$,
$\pi \Delta$, $\eta N$, $K \Lambda$ and $K \Sigma$. Our goal is to
determine the vector-meson--nucleon scattering amplitude and then
compute the self energy of a vector meson in nuclear matter to
leading order in density. In this work we focus on vector mesons
with small or zero 3-momentum with respect to the nuclear medium.
Therefore it is sufficient to consider only s-wave scattering in
the $\rho N$ and $\omega N$ channels~\footnote{Also in the other
`heavy' channels ($\eta N, K \Lambda, K \Sigma$) only s-waves are
considered so far.}. This implies that in the $\pi N$ channel we
need to consider only the $S_{11}, S_{31}, D_{13}$ and $D_{33}$
partial waves. Furthermore, we include the pion-induced production
of $\eta$, $\omega$ and $\rho$ mesons off nucleons as well as the
reactions $\pi^- p \rightarrow K^0 \Lambda$ and $\pi^+ p
\rightarrow K^+ \Sigma^+$. In addition to the hadronic observables
employed in the fit, we use photo-induced production of
pseudoscalar mesons, assuming vector-meson dominance, to eliminate
an ambiguity in the $S_{11}$ channel.

\begin{figure}[t]
\setlength{\unitlength}{1mm}
\begin{picture}(140,65)
\put(68,0){\epsfig{file=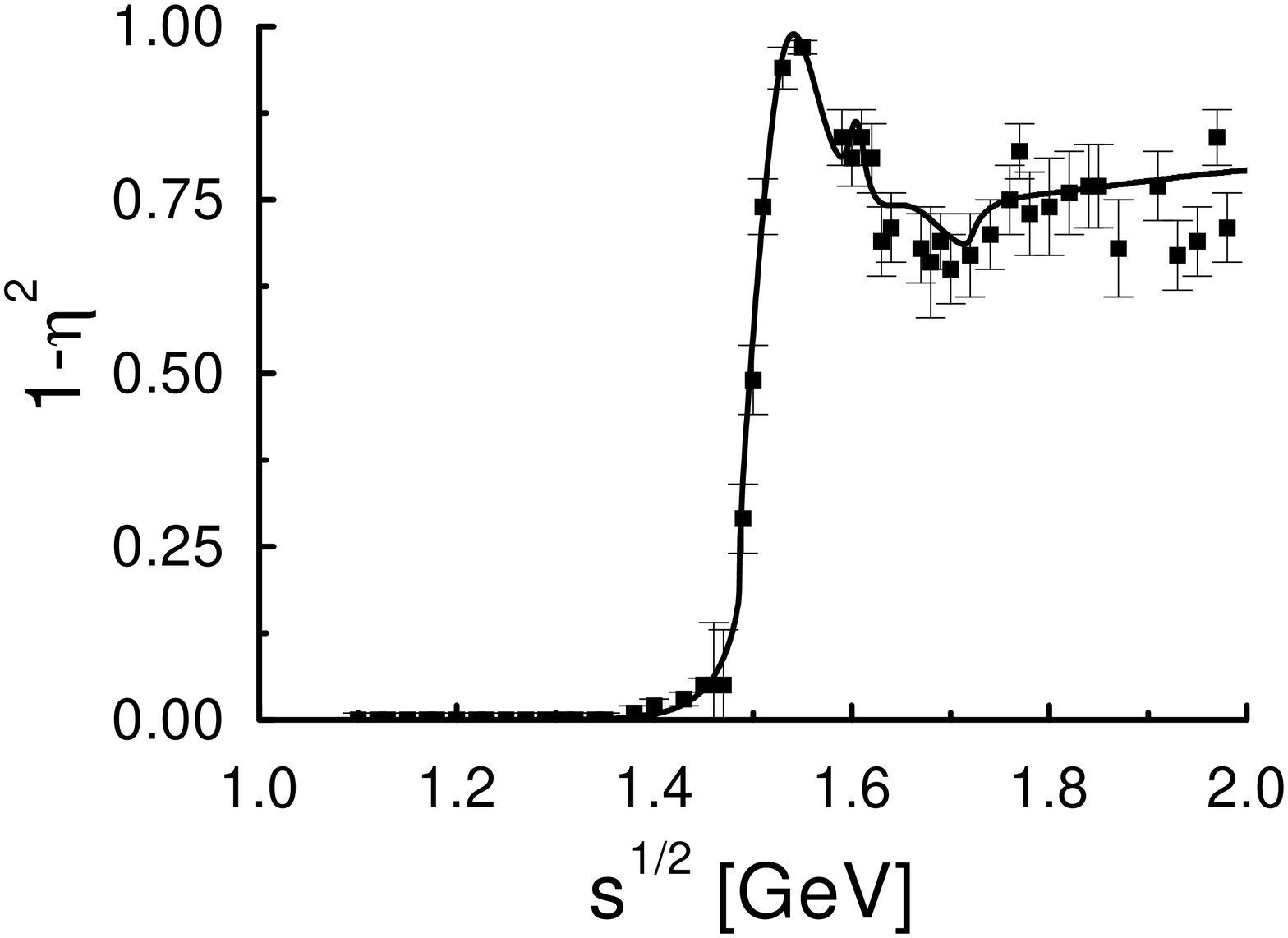,height=53mm}}
\put(0,0){\epsfig{file=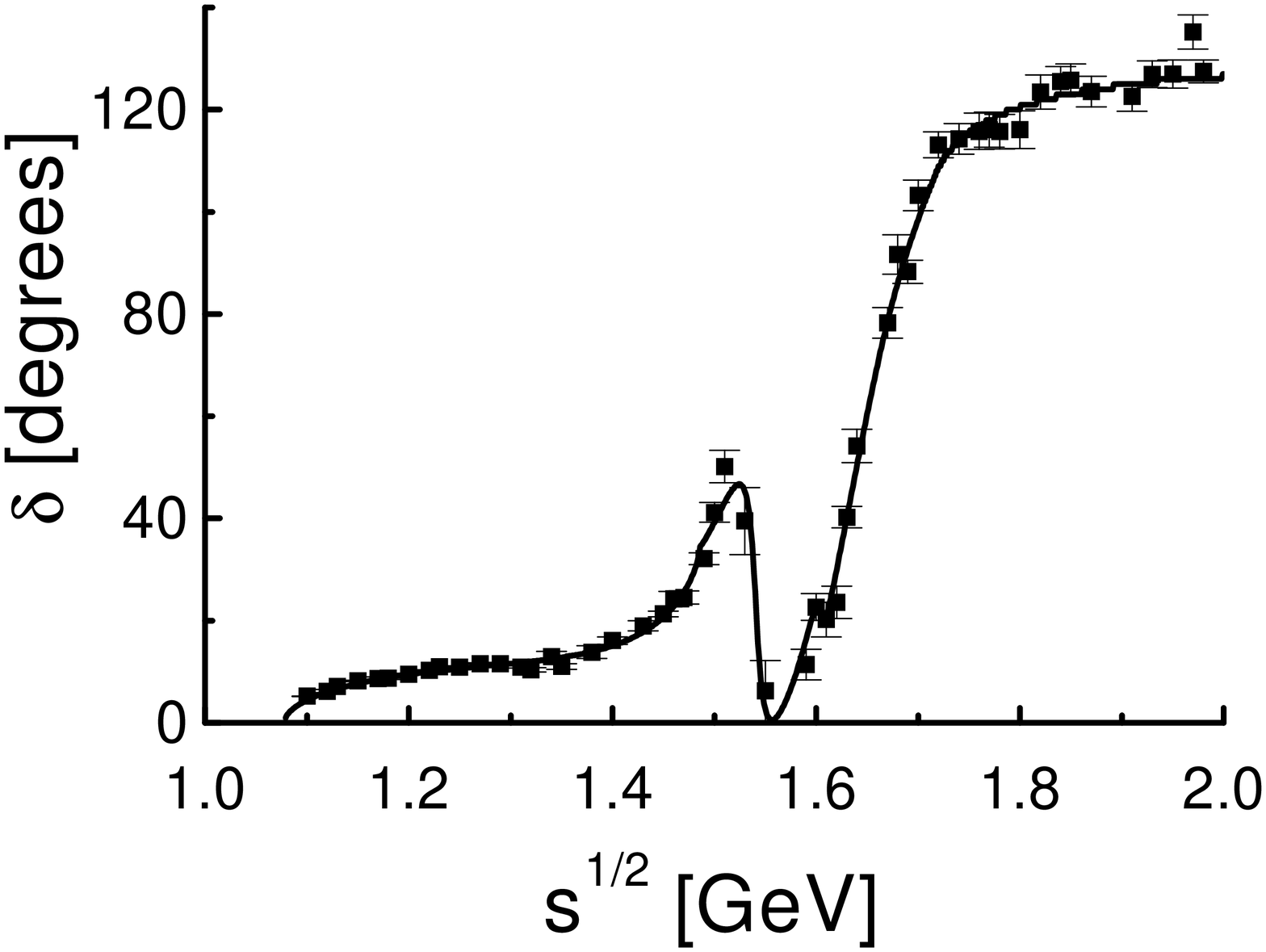,height=53mm}}
\end{picture}
\caption{\label{S11-scatt} The $\pi N$ scattering phase shifts
and inelasticity in the $S_{11}$ channel. The line shows the best
fit, while the data points are from the analysis of Arndt {\em et
al.}~$[11]$.}
\end{figure}

We recall that at small nuclear matter densities the spectral
function of a vector meson with energy $\omega $ and zero momentum
probes the vector-meson nucleon scattering  process at $\sqrt{s}
\sim m_N+\omega$. In order to learn something about the momentum
dependence of the vector-meson self energy, vector-meson--nucleon
scattering also in higher partial waves would have to be
considered~\cite{FP}.

In accordance with the ideas outlined above, only data in the
relevant kinematical range will be used in the analysis. The
threshold for vector-meson production off a nucleon is at $\sqrt{s}
\simeq 1.7$ GeV. We fit the data in the energy window~\footnote{In
the s-wave $\pi N$ scattering channels we extend the energy window
down to $1.2$ GeV.} $1.4$ GeV $\leq\sqrt{s}
\leq 1.9$ GeV, using an effective Lagrangian with local 4-point
meson-meson--baryon-baryon interactions. For details the reader is
referred to ref.~\cite{LWF}.

\begin{figure}[t]
\setlength{\unitlength}{1mm}
\begin{picture}(140,110)
\put(70,50){\epsfig{file=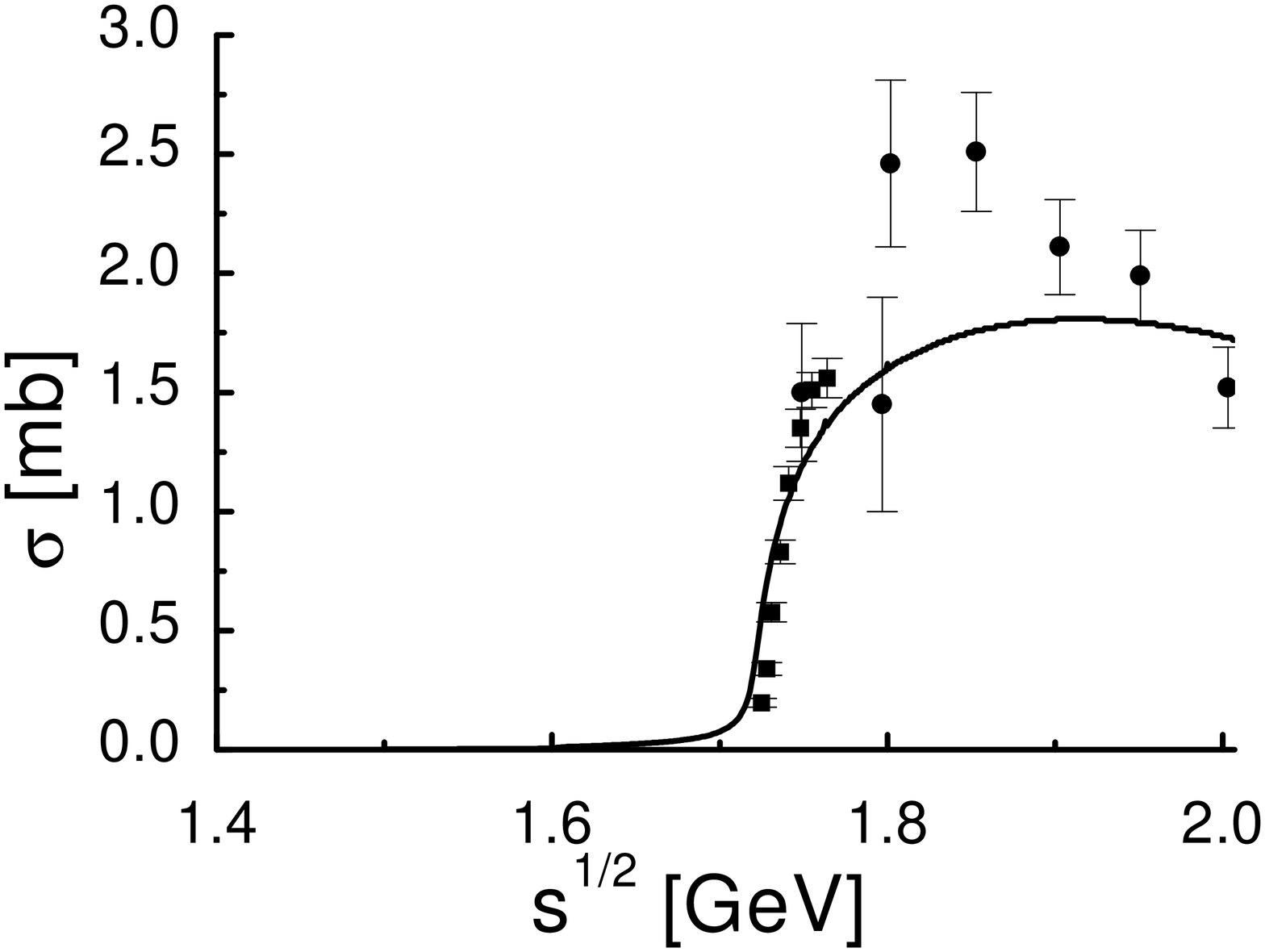,height=54mm}}
\put(0,50){\epsfig{file=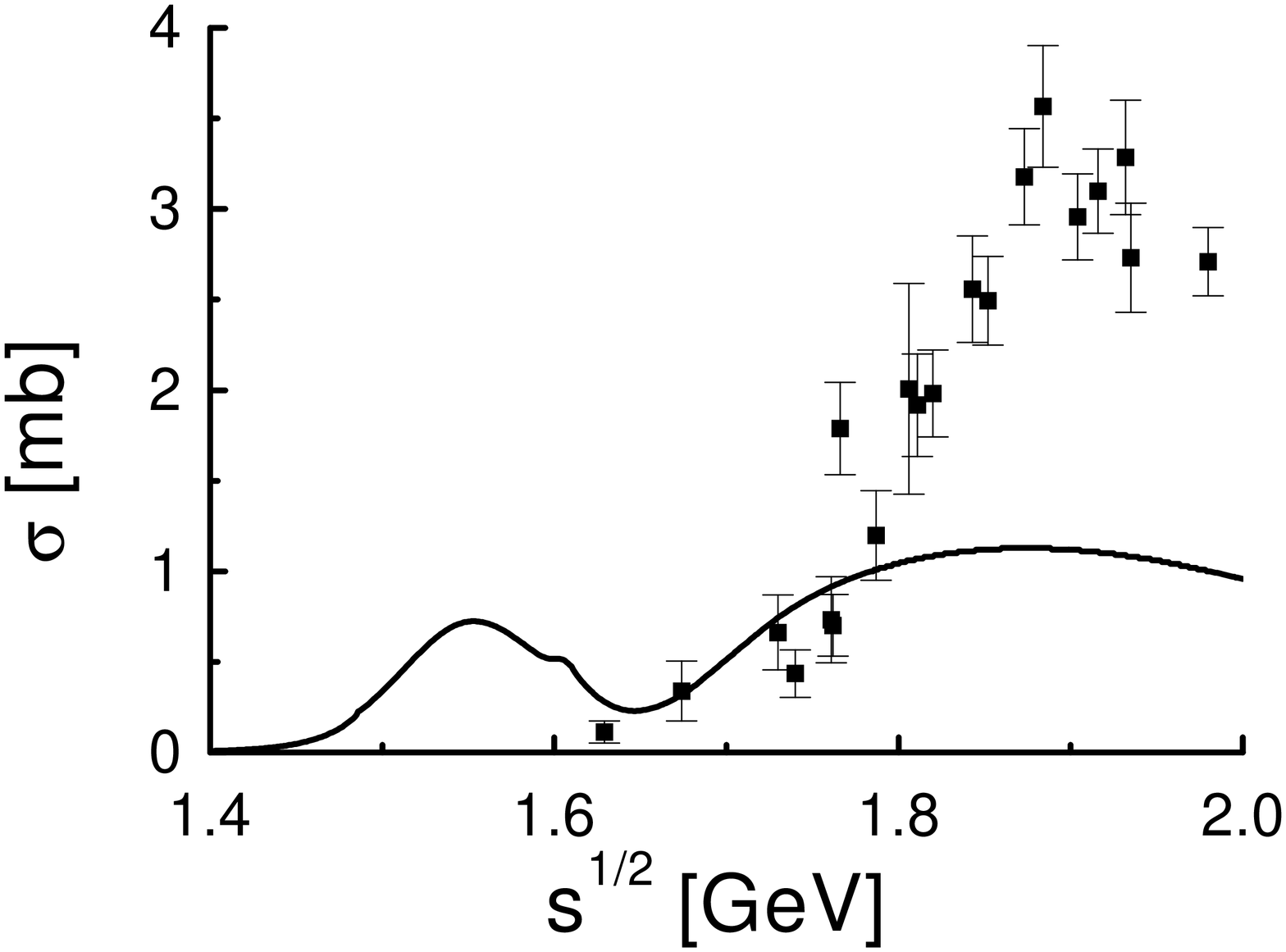,height=54mm}}
\put(25,90){a)}
\put(95,90){b)}
\put(2,0){\epsfig{file=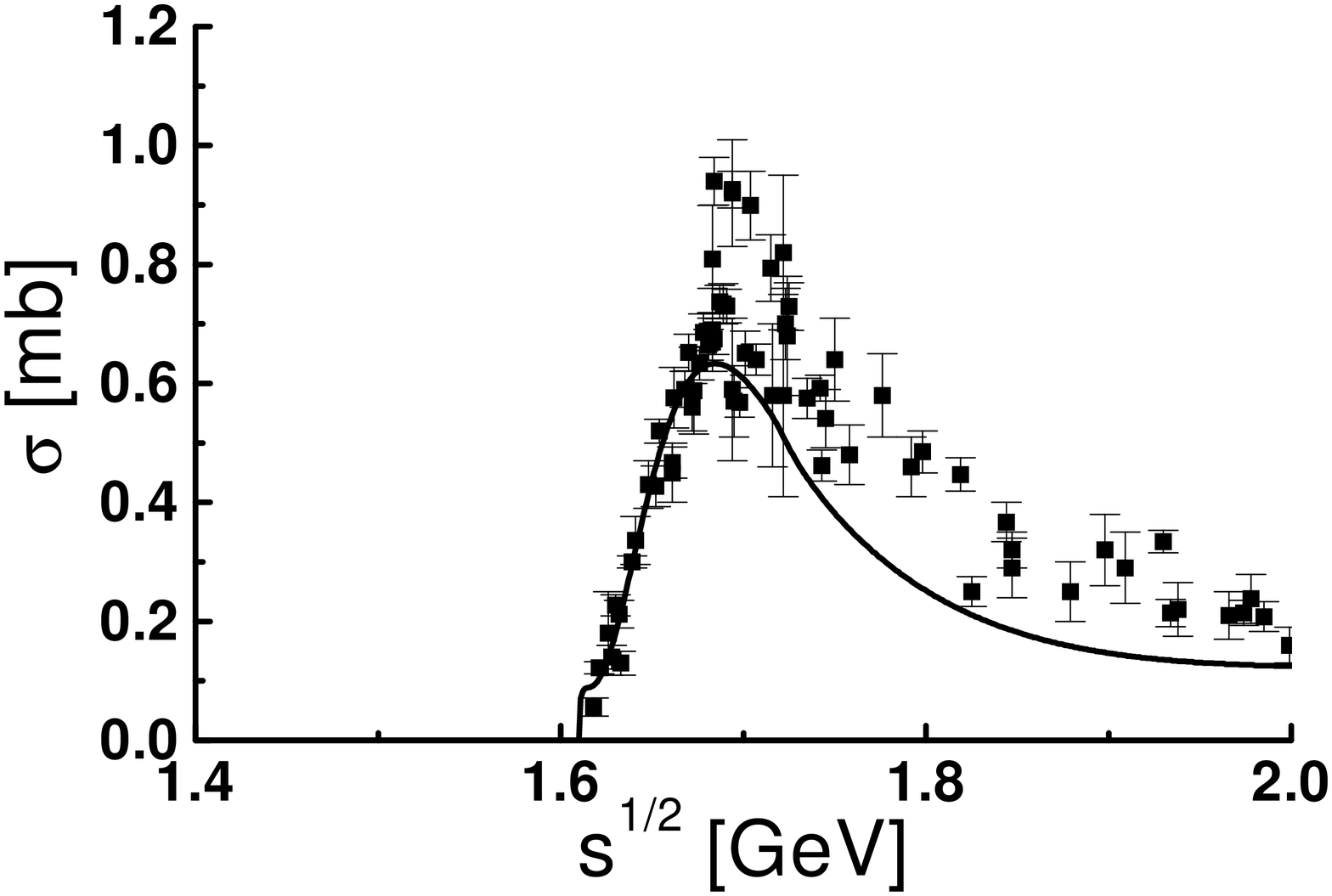,height=50mm}}
\put(25,35){c)}
\put(72,0){\epsfig{file=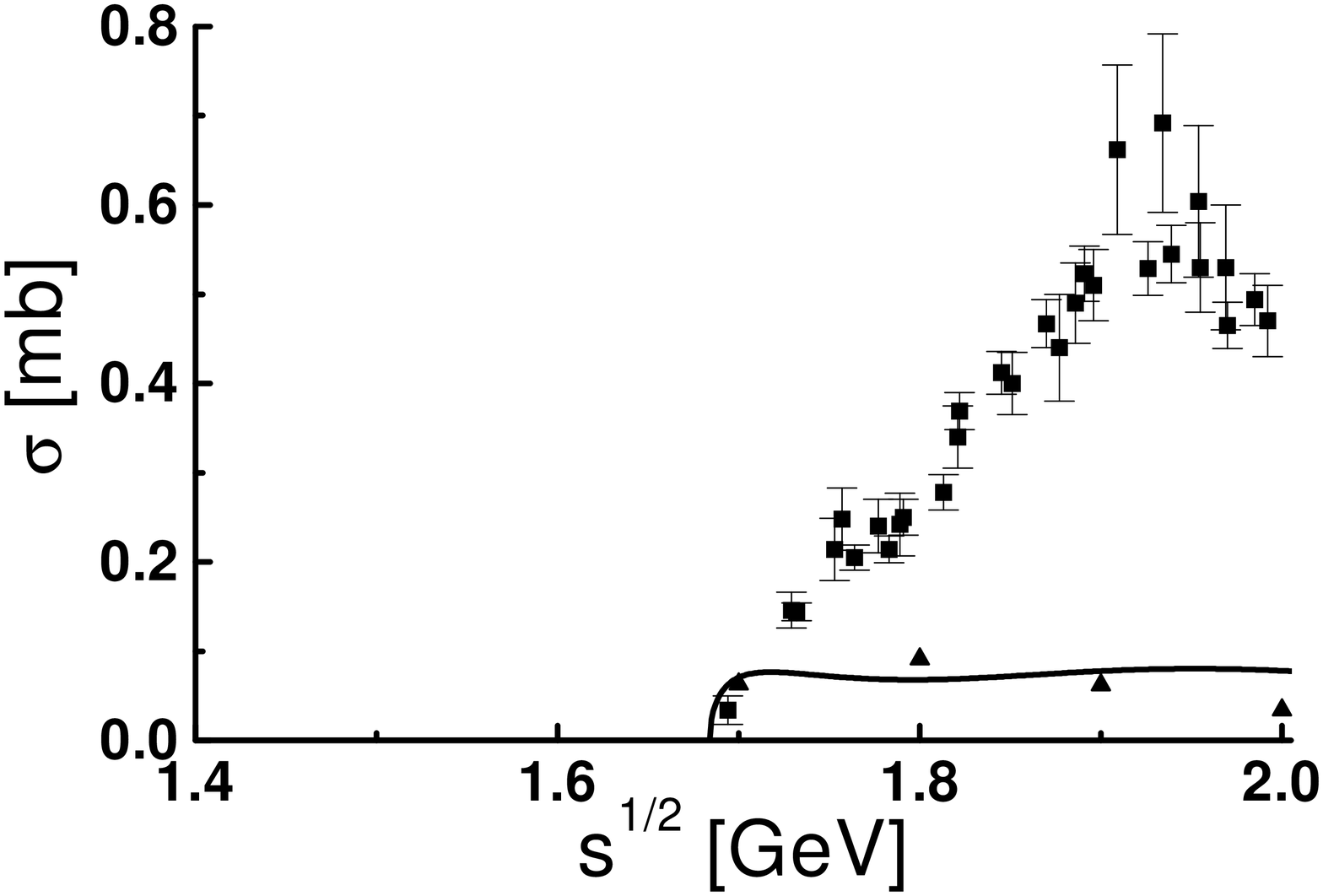,height=50mm}}
\put(95,35){d)}
\end{picture}
\caption{\label{rho-prod} The cross sections for the reactions
a) $\pi^- p \rightarrow \rho^0 n$, b) $\pi^- p \rightarrow \omega
n$, c) $\pi^- p\rightarrow K^0 \Lambda$ and d) $\pi^+ p\rightarrow
K^+ \Sigma^+$. The data are from ref.~$[12]$ and the partial-wave
analysis used in d) from ref.~$[13]$.}
\end{figure}

In fig.~\ref{S11-scatt} our result for the $\pi N$ scattering is
illustrated by the $S_{11}$ channel. In the remaining channels the
fit is of similar quality. Furthermore, in fig.~\ref{rho-prod} the
cross sections for the reactions $\pi^- p\rightarrow \rho^0 n$,
$\pi^- p\rightarrow
\omega n$,  $\pi^- p\rightarrow K^0 \Lambda$ and
$\pi^+ p\rightarrow K^+ \Sigma^+$ are shown. The agreement with the
data is good close to threshold, where s-waves in the final state
are expected to dominate. At higher energies there is room for
higher partial waves. For the $K^+\Sigma^+$ channel we use the
partial-wave analysis of ref.~\cite{candlin} to extract the s-wave
contribution to the cross section, shown by the triangles in
fig.~\ref{rho-prod} d. Not shown are the cross sections for the
reactions $\pi^- p\rightarrow \eta n$ and $\pi^+ p\rightarrow
\rho^+ p$. We obtain a good description of the former
up to $s^{1/2} \simeq 1.7$ GeV, where the p-wave contribution is
expected to set in. In view of the scarce data at low energies in
the latter channel, we find a reasonable fit~\cite{LWF}.

The bumps in the $\rho$-production cross section at $\sqrt{s}$
below 1.6 GeV are due to the coupling to resonances below the
threshold, like the $N^\star(1520)$. This indicates that these
resonances play an important role in the $\rho$-nucleon dynamics,
in agreement with the results of Manley and Saleski~\cite{manley}.
We have extracted coupling constants for the vector mesons to some
of the resonances. For instance, for the $N^\star (1520)$ we use
the non-relativistic interaction Lagrangian~\cite{peters}
\begin{equation}
{\mathcal L}_{int}= \frac{f_{N^\star N
V}}{m_V}\Psi^\dagger_{N^\star} (\vec{S}\cdot \vec{V} q_0 - V_0
\vec{S}\cdot \vec{q})\Psi_N + h.c.\;,
\end{equation}
where $\vec{S}$ is the transition spin operator and $V^\mu = {\bf
\rho}^\mu\cdot{\bf\tau},\omega^\mu$. The $\rho N \rightarrow \rho N$ and $\omega N
\rightarrow \omega N$ scattering amplitudes obtained in our model
are then fitted with Breit-Wigner amplitudes, where the decay of
the resonance into the vector-meson--nucleon channel is described
by the interaction (1). We find a $\rho$ meson coupling to the
$N^\star (1520)$ of $f_{N^\star N \rho} =  3.2$.  Note that our
value is a factor two smaller than that extracted by Peters {\em et
al.}~\cite{peters}. This is at least in part due to the fact that
we use the cross section for pion-induced $\rho$ production of
Brody {\em et. al}~\cite{brody} (Fig.~\ref{rho-prod} a)), while
Peters {\em et al.}~\cite{peters} use a partial width for the decay
$N^\star(1520)\rightarrow \rho N$, extracted from the analysis of
Manley and Saleski~\cite{manley}. Close to the
threshold~\cite{post}, the cross section implicit in the analysis
of Manley and Saleski is much larger than that of Brody {\em et.
al}. This difference may reflect the difficulties involved in
extracting the $\rho$ production cross section close to threshold.
Data on pion-induced production of $e^+ e^-$ pairs in this energy
regime would provide additional constraints on the amplitudes,
which may allow one to greatly reduce this
uncertainty~\cite{Madeleine}.

We also find a strong $\omega$ coupling to the $N^\star(1520)$,
$f_{N^\star N \omega}= 6.5$. Furthermore, we find that the $N^\star
(1535)$ resonance couples strongly to the $\rho$ channel, while the
$N^\star (1650)$ interacts strongly with both vector-meson
channels. These findings are qualitatively consistent with the
photon decay helicity amplitudes, assuming vector-meson
dominance~\cite{post,PDG}. We note the the hadronic observables
used in the fit do not determine the relative phases of
off-diagonal amplitudes, like e.g. those of the $\pi N
\rightarrow \rho N$ and $\pi N \rightarrow \omega N$ reactions.
These phases are of course cruical for the interference in the
pion-induced production of $e^+ e^-$ pairs~\cite{Madeleine}. We
determine the phases by comparing with the photon-decay helicity
amplitudes of the resonances in each channel~\cite{PDG}.

\begin{figure}[t]
\setlength{\unitlength}{1mm}
\begin{picture}(140,60)
\put(0,0){\epsfig{file=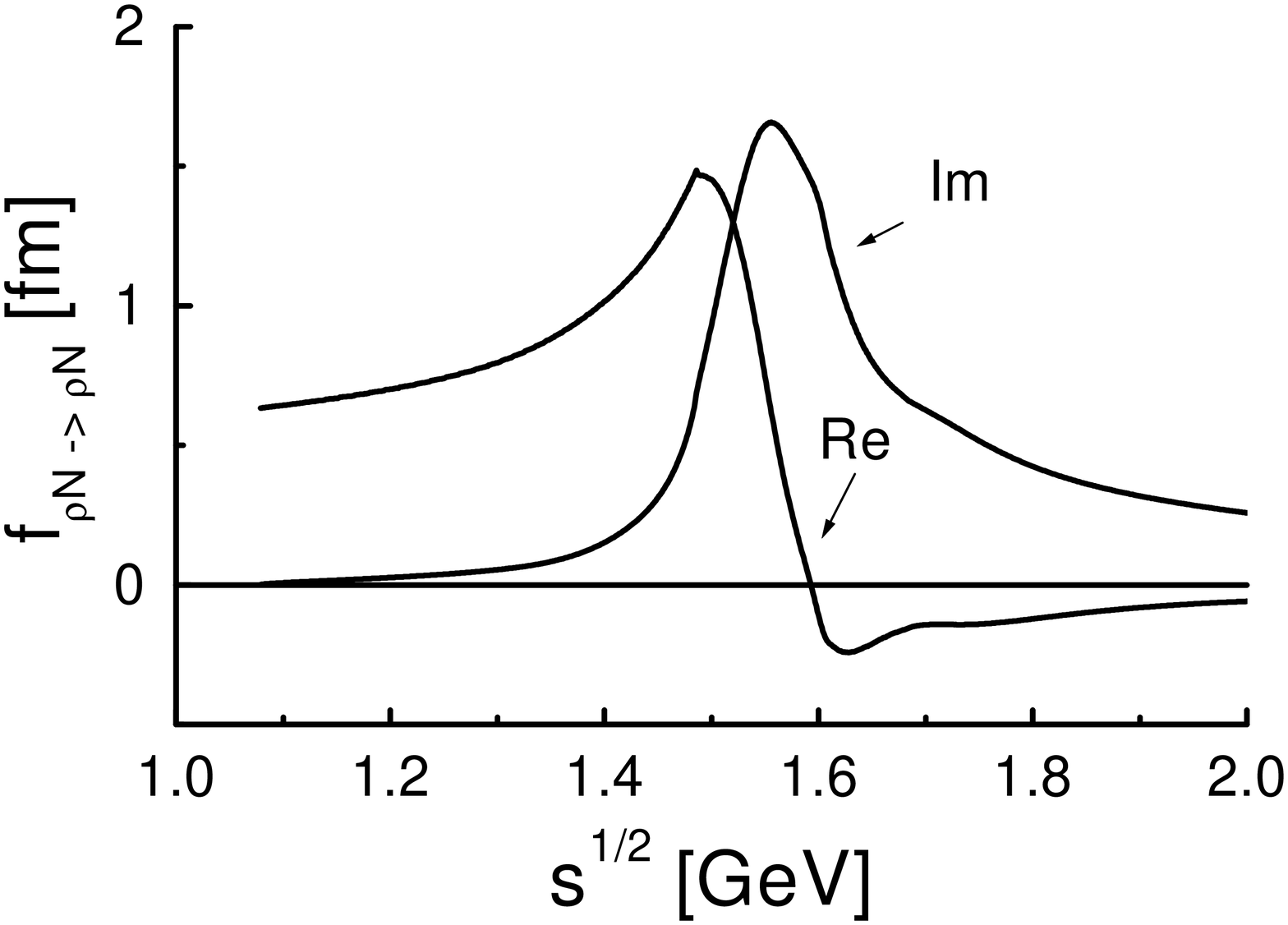,height=50mm}}
\put(72,1.2){\epsfig{file=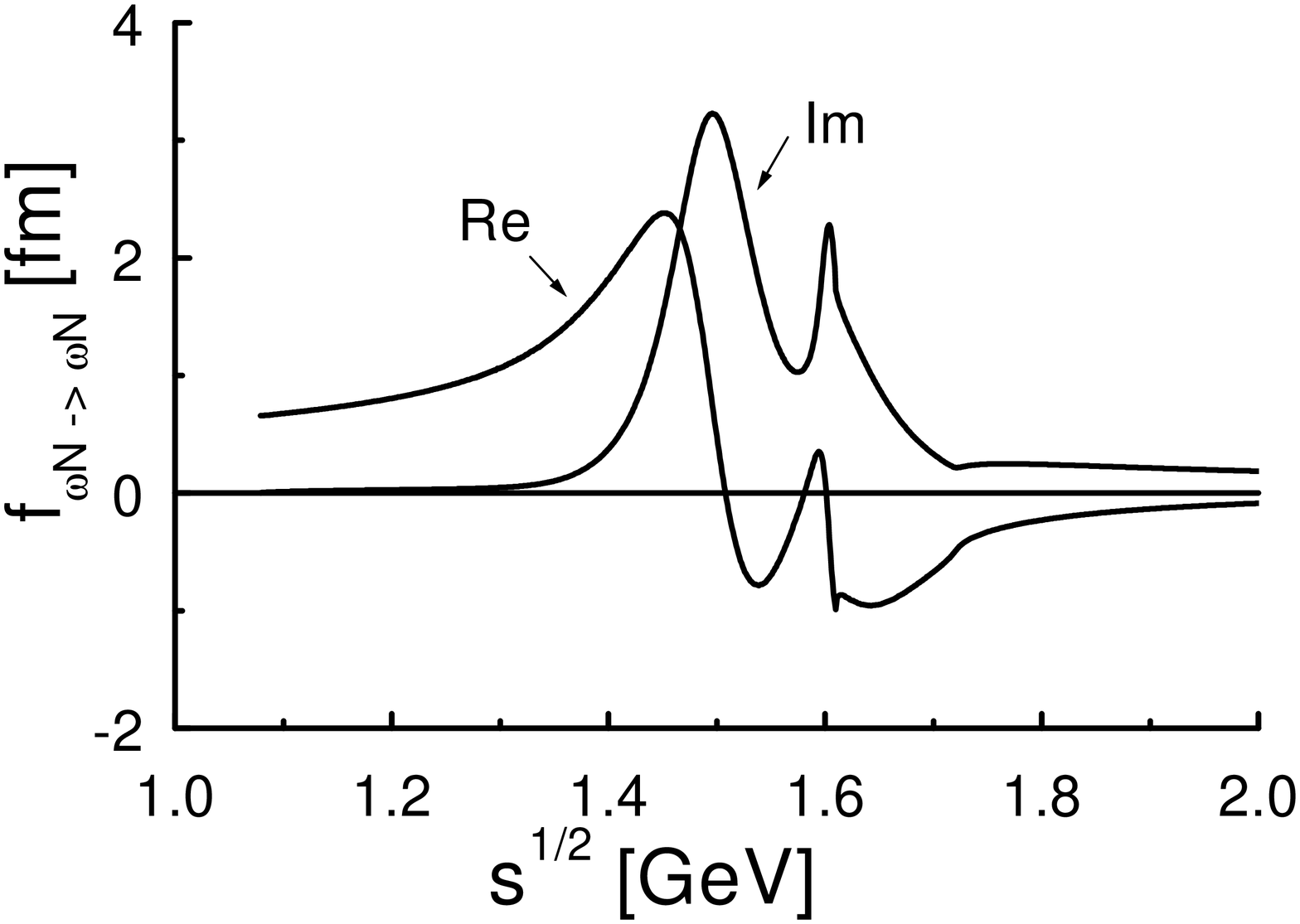,height=50mm}}
\end{picture}
\caption{\label{amplitudes} The $\rho N$ and $\omega N$
scattering amplitudes, averaged over spin and isospin.}
\end{figure}
The resulting $\rho$- and $\omega$-nucleon scattering amplitudes
are shown in Fig.~\ref{amplitudes}. The $\rho-N$ and $\omega-N$
scattering lengths, defined by $a_{VN}=f_{VN}(\sqrt{s}=m_N+m_V)$,
are $a_{\rho N}=$ (-0.1+0.6\,i) fm and $a_{\omega N}=$
(-0.5+0.2\,i) fm. To lowest order in density, this corresponds to
the following in-medium modifications of masses and widths at
nuclear matter density: $\Delta m_\rho
\simeq 10$ MeV, $\Delta m_\omega
\simeq 50$ MeV, $\Delta\Gamma_\rho \simeq 120$ MeV and $\Delta
\Gamma_\omega \simeq 40$ MeV. However, as we show in the next
section, the coupling of the vector mesons to baryon resonances
below threshold, which is reflected in the strong energy dependence
of the amplitudes, cannot be neglected.

\section{Vector mesons in nuclear matter}

In this section we present results for the propagators of the
$\rho$- and $\omega$-mesons at rest in nuclear matter, obtained
with the scattering amplitudes presented in section 2, to leading
order in density. The low-density theorem states that the self
energy, $\Delta m_V^2(\omega )$, of a vector meson $V$ in nuclear
matter is given by~\cite{LDT}
\begin{equation}
\Delta m_V^2(\omega ) = -4\pi(1+\frac{\omega}{m_N})
f_{V N}\,(\sqrt{s}=m_N+\omega ) \rho_N + \dots,
\label{LDT}
\end{equation}
\begin{figure}[t]
\setlength{\unitlength}{1mm}
\begin{picture}(140,60)
\put(1,1.5){\epsfig{file=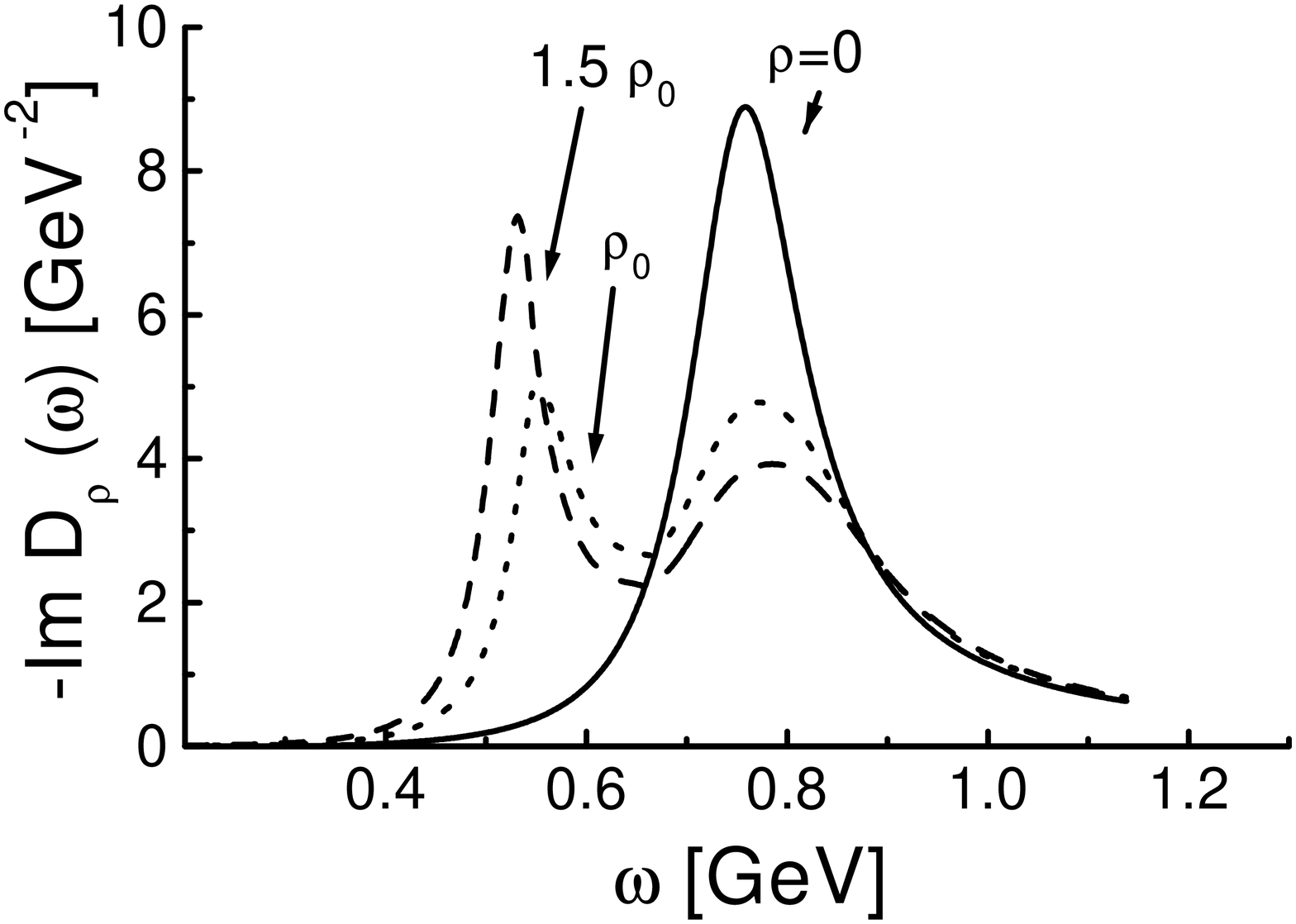,height=52mm}}
\put(72,0){\epsfig{file=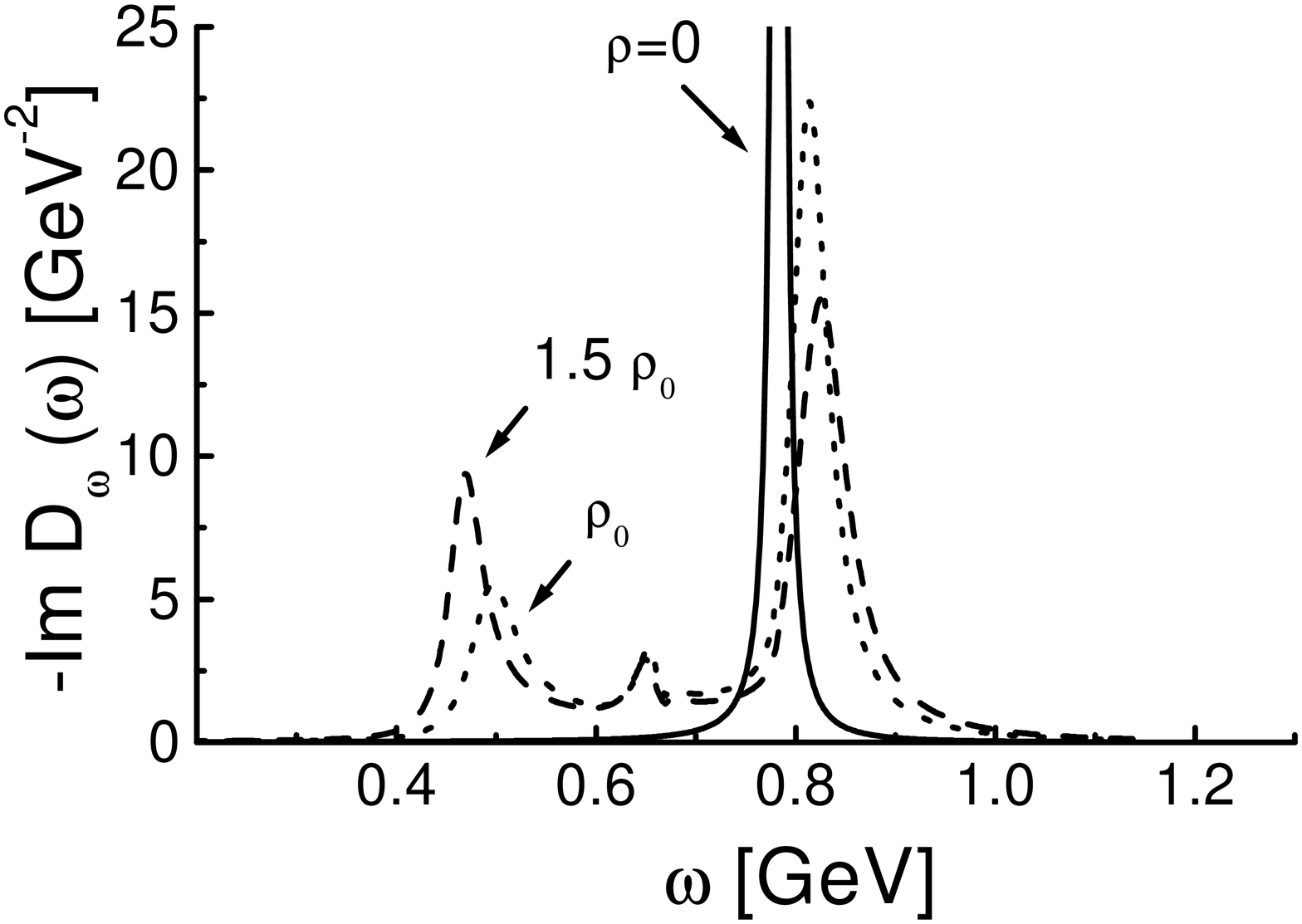,height=52mm}}
\end{picture}
\caption{\label{propagators} Imaginary parts of the
$\rho$ and $\omega$ propagators in nuclear matter at $\rho
= \rho_0$ and $1.5 \rho_0$, compared those in vacuum.}
\end{figure}
where $\omega$ is the energy of the vector meson, $m_N$ the nucleon
mass, $\rho_N$ the nucleon density and ${f}_{V N}$ denotes the $V
N$ s-wave scattering amplitude averaged over spin and isospin. In
fig.~\ref{propagators} we show the resulting propagators at the
saturation density of nuclear matter, $\rho_0 = 0.17
\mbox{~fm}^{-3}$ and at $\rho = 1.5 \rho_0$. For the $\rho$ meson
we note a strong enhancement of the width, and a downward shift in
energy, due to the mixing with the baryon resonances at $\sqrt{s} =
1.5 - 1.6$ GeV. At $\rho=\rho_0$ the lower peak carries about
$20$~\% of the energy-weighted sum rule, while the
center-of-gravity of the spectral function is shifted down in
energy by $\simeq 10$~\%. As the density is increased, more
strength is shifted down to the resonance-hole like peak at low
energies and the width of the $\rho$-like peak is enhanced.

The in-medium propagator of the $\omega$ meson exhibits three
distinct quasiparticles, an $\omega$ like mode, which is shifted up
somewhat in energy, and resonance-hole like modes at low energies.
The low-lying modes carry about 20 \% on the energy-weighted sum
rule. Again, the center-of-gravity is shifted down by $\simeq
10$~\%. However, we stress that the structure of the in-medium
$\omega$ spectral function clearly cannot be characterized by this
number alone.

We expect that the results obtained with only the leading term in
the low-density expansion are qualitatively correct at normal
nuclear matter density. However, on a quantitative level, the
spectral functions may change when higher order terms in the
density expansion  are included. For instance, it may be that the
in-medium properties of the baryon resonances depend sensitively on
the meson spectral functions. If this is the case, a self
consistent calculation, which corresponds to a partial summation of
terms in the density expansion, would have to be
performed~\cite{lutz,peters}.

\begin{figure}[t]
\setlength{\unitlength}{1mm}
\begin{picture}(140,60)
\put(0,0){\epsfig{file=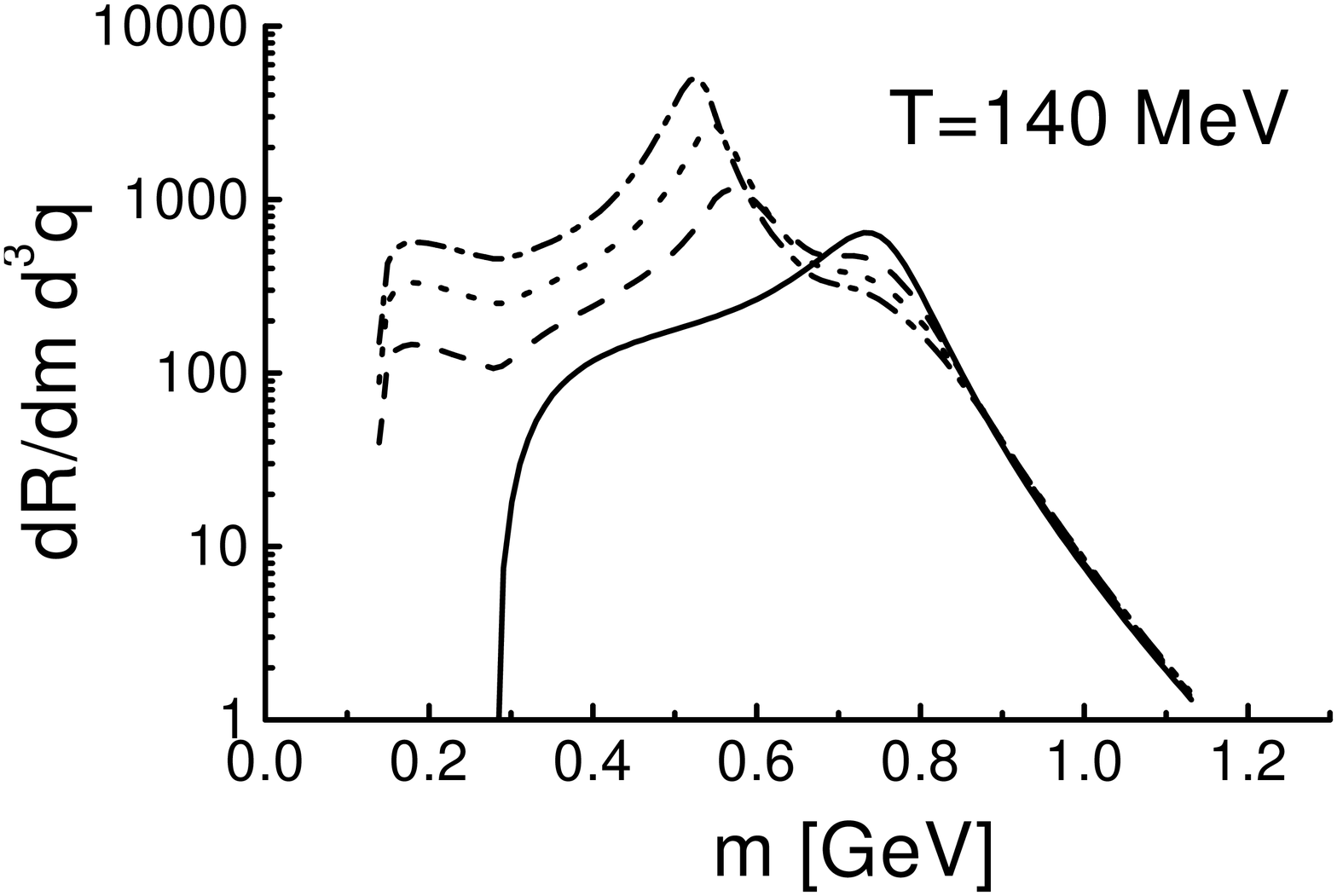,height=50mm}}
\put(71,0){\epsfig{file=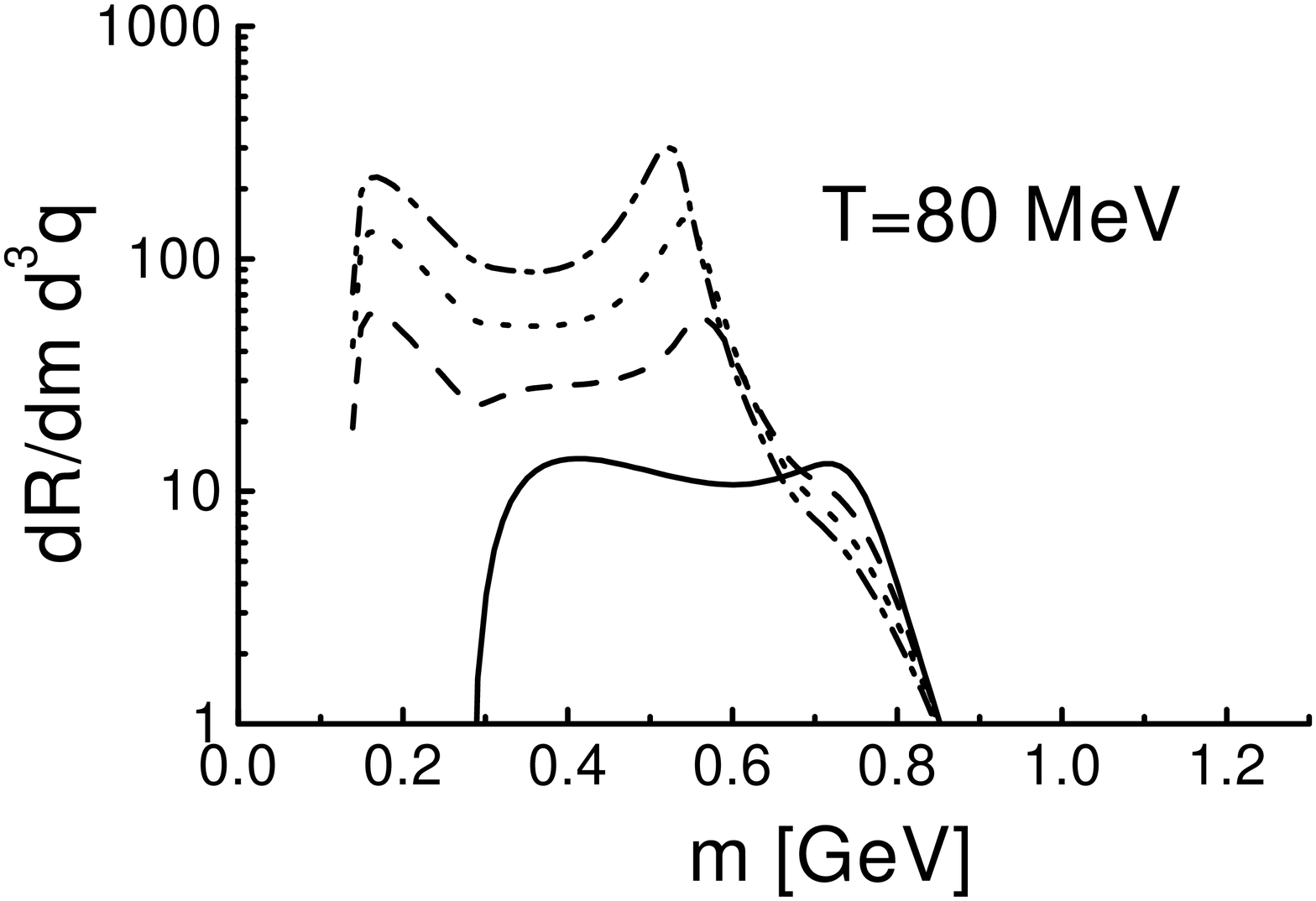,height=50mm}}
\end{picture}
\caption{\label{lepton-pairs} The thermal production rate for
back-to-back lepton-pairs at T$=80$ and 140 and $\rho
= 0.5$ (dashed line), $1.0$ (dotted line) and $1.5 \rho_0$
(dash-dotted line). }
\end{figure}

Finally, in Fig.~\ref{lepton-pairs} we show thermal rates for
lepton-pair production due to the decay of in-medium $\rho$ and
$\omega$ mesons in uniform nuclear matter at temperatures $T=80$
and 140 MeV and densities $\rho/\rho_0 = 0, 0.5, 1.0, 1.5$. We
consider only back-to-back pairs, since the model includes only
s-wave vector-meson-nucleon interactions so far. At $T = 140$ MeV,
which corresponds roughly to the conditions expected in heavy-ion
collisions at CERN energies, the resonance-hole like mode in the
$\rho$ channel gives rise to a strong enhancement in the
invariant-mass region 400-600 MeV, where the CERES and HELIOS-3
experiments find an excess of lepton
pairs~\cite{CERES,misko,Helios3}. The low-energy tail of the
$\rho$-meson spectral function yields a fairly constant
contribution down to the pion threshold. At the lower temperature
($T = 80$ MeV), which approximately corresponds to GSI/SIS
energies, the relative enhancement is even stronger.

\section{Conclusions}

A relativistic and unitary, coupled channel approach to
meson-nucleon scattering was presented. The parameters of the
effective interaction are determined by fitting elastic
pion-nucleon scattering and pion-induced meson production data in
the relevant energy regime. We obtain a good overall description of
the data and a model for the $\rho$- and $\omega$-N scattering
amplitudes, which in turn allows us to compute the vector-meson
self energies in nuclear matter. In this approach we minimize the
model dependence and avoid the potentially dangerous extrapolation
from low-energy data~\cite{FP,bfseoul}. Further constraints on the
vector-meson scattering amplitudes, in particular in the
subthreshold regime, can be obtained from the reaction $\pi N
\rightarrow e^+ e^- N$, which is discussed in the talk of M. Soyeur
\cite{Madeleine}.

An prominent feature of the scattering amplitudes is the strong
coupling to baryon resonances below threshold. This leads to two
characteristic features of the vector-meson spectral functions,
namely repulsive scattering lengths and a spreading of the
vector-meson strength to states at low energy. The latter is
qualitatively what seems to be required by the heavy-ion data.
Clearly, complementary experiments with e.g. photon and pion
induced vector meson production off nuclei would be extremely
useful for exploring the in-medium properties of these mesons in
more detail.



\begin{thebibliography}{99}
\bibitem{CERES} CERES, G.~Agakichiev {\em et al.},
\Journal{\PRL}{75}{1272}{1995}; \Journal{\PLB}{422}{405}{1998};
\Journal{\NPA}{661}{23c}{1999}
\bibitem{misko} D.~Miskowiec, these proceedings
\bibitem{Helios3} HELIOS-3, M. Masera {\em et al.},
\Journal{\NPA}{590}{93c}{1992}
\bibitem{Drees} A.~Drees, \Journal{\NPA}{610}{536c}{1996}
\bibitem{likobrown} G.Q.~Li, C.M.~Ko and G.E.~Brown,
\Journal{\PRL}{75}{4007}{1995}
\bibitem{BR} G.E.~Brown and M.~Rho, \Journal{\PRL}{66}{2720}{1991}
\bibitem{RCW} R.~Rapp, G.~Chanfray, J.~Wambach,
\Journal{\NPA}{617}{472}{1997}
\bibitem{LDT} W.~Lenz, \Journal{\ZP}{56}{778}{1929};
C.D.~Dover, J.~H\"ufner and R.H.~Lemmer,
\Journal{\AP}{66}{248}{1971}; M.~Lutz, A.~Steiner and W.~Weise,
\Journal{\NPA}{574}{755}{1994}
\bibitem{LWF} M.~Lutz, G.~Wolf and B.~Friman, to be published
\bibitem{FP} B.~Friman and H.J.~Pirner, \Journal{\NPA}{617}{496}{1997}
\bibitem{arndt} R.A.~Arndt {\em et al.}, \Journal{\PRC}{52}{2120}{1995}
\bibitem{LB} A.~Baldini {\em et al.}, in Landolt-B\"ornstein,
vol 1/12a (Springer, Berlin, 1987)
\bibitem{candlin} D.J.~Candlin {\em et al.} \Journal{\NPB}{238}{477}{1984}
\bibitem{manley} D.M.~Manley and E.M.~Saleski, \Journal{\PRC}{45}{4002}{1992}
\bibitem{peters} W.~Peters, M.~Post, H.~Lenske, S. Leupold and U.~Mosel,
\Journal{\NPA}{632}{109}{1998}
\bibitem{brody} A.D.~Brody {\em et al.}, \Journal{\PRD}{4}{2693}{1971}
\bibitem{post} M.~Post, private communication
\bibitem{Madeleine} M.~Soyeur, M.~Lutz and B.~Friman, these proceedings
\bibitem{PDG} Particle Data Group, Eur. Phys. J. {\bf C 3} (1998) 1
\bibitem{lutz} M.~Lutz, \Journal{\PLB}{426}{12}{1998}
\bibitem{bfseoul} B.~Friman, in Proc. Workshop on
Astro-Hadron Physics, Seoul, Korea, Oct., 1997, nucl-th/9801053


\end{thebibliography}
\end{document}